\documentclass[twocolumn,showpacs,superscriptaddress,amsmath,amssymb,aps,prb]{revtex4}
\pdfoutput=1

\usepackage{graphicx}
\usepackage{dcolumn}
\usepackage{bm}

\begin{document}

\preprint{APS/123-QED}

\title{Controllable spin singlet - spin triplet transition in three concentric quantum rings through magnetic field and confinement potential}

\author{Bin Li}
 \email{phymilky@gmail.com}
 \affiliation{Departement Fysica, Universiteit Antwerpen, Groenenborgerlaan 171, B-2020 Antwerpen, Belgium.}
 \author{F. M. Peeters}
 \email{Francois.Peeters@ua.ac.be}
 \affiliation{Departement Fysica, Universiteit Antwerpen, Groenenborgerlaan 171, B-2020 Antwerpen, Belgium.}

\date{\today}

\begin{abstract}
We present a theoretical study of the spectrum of electrons confined in triple concentric rings. An unusual ordering and rich variety of angular momentum transitions are found that depend on the coupling between the rings and the confinement potential of the rings. Using the Configuration Interaction (CI) method, we calculated the two electron energy spectrum. Spin singlet to spin triplet transitions of the electron ground state are predicted and a fractional Aharonov-Bohm effect is found. We show that both the period and amplitude of the spin singlet - triplet energy gap depend strongly on the confinement potential and the external magnetic field. The spin singlet - triplet transition is found to depend on the spin Zeeman energy, especially for rings with weak confinement and in the presence of large magnetic field. The amplitude of the spin singlet - triplet energy gap depends on the Land\'{e} $g$-factor but the period of the transitions is independent of $g$.
\end{abstract}

\maketitle

\section{\label{sec:introduction}Introduction}
Spin qubits~\cite{Loss1998,Barenco1995,Brum1997,Kane1998}, whose dephasing time can be on the order of microseconds, thus several orders longer than those of charge qubits, are promising proposals for scalable solid-state quantum bits. Spin qubits based quantum gate mechanism was first proposed by Loss and Divincenzo~\cite{Loss1998}. Later, Burkard and his collaborators~\cite{Burkard1999} showed that the ground state spin configuration of two electrons in coupled quantum dots can switch between a spin-singlet ($S=0$) and a spin-triplet state ($S=1$). The manipulation of the switching between the two spin states can be realized by changing the applied gate voltage in the presence of a finite applied magnetic field~\cite{Kyriakidis2002,Lee2004}. Such spin qubits can also be realized with two electron concentric quantum rings. As compared to quantum dots, quantum rings have a very different topology due to the presence of the central hole. As a result, the single particle energy levels of quantum rings have a unique magnetic field dispersion, giving rise to the Aharonov-Bohm (AB) effect where the quantized ground state total angular momentum increases with magnetic field~\cite{Fuhrer2001,Lorke2000}. Since in a multi-ring configuration the electrons may switch between different rings, resulting in different patterns of AB oscillations in the same structure~\cite{Szafran2005}. A few electron system in such structures may exhibit a fractional Aharonov-Bohm effect~\cite{Chakraborty1994,Niemela1996} due to the correlations~\cite{Niemela1996} between the electrons.

Quantum rings can be realized by using local oxidation techniques (lithography)~\cite{Fuhrer2001,Wiel2003a}, as well as self-assembly with~\cite{Garcia1997} or without~\cite{Gong2005,Mano2005,Kuroda2005} a capping layer, such as droplet-epitaxial growth techniques~\cite{Mano2005,Somaschini2009}. By changing the substrate temperature during the quantum ring growth, it is possible to fabricate not only concentric double quantum rings (CDQRs), but also multiple concentric quantum rings (MCQRs)~\cite{Somaschini2009,Somaschini2010}. The radii of the rings, together with the confinement in each ring, may strongly affect the spectrum~\cite{Szafran2005,Climente2006,Planelles2005}. The magnetic dependence of the two and few electron spectra in a single quantum ring\cite{Chakraborty1994,Niemela1996}, CDQRS~\cite{Szafran2005,Climente2006,Escartin2009} and MCQRs\cite{Escartin2010} has already been studied theoretically. Previous studies showed that transitions between singlet and triplet spin states in quantum dots can be controlled by the applied gate voltage which controls the confinement\cite{Wagner1992,Kyriakidis2002}. Earlier in 2005, Szafran and Peeters~\cite{Szafran2005} studied the few electron eigenstates in CDQRs in case of different inter-ring coupling (in their paper, it depends on the distance between the rings), while they did not focus on the details of the confinement (or coupling) dependent spin-singlet spin-triplet transition. Later in 2010, Escart\'{\i}n reported interesting few electrons ground state and persistent current results in triple concentric quantum rings with strong confinement\cite{Escartin2010}. However, studies with full detail of the spin singlet spin triplet transition and their dependence on an external magnetic field or confinement is still lacking for MCQRs, while it is essential for their eventual application in a practical implementation for quantum information. Such calculations will be presented in this paper. We will show an unusual ordering of angular momentum transitions in MCQRs that strongly depend on the confinement, and we studied the period and amplitude of the spin singlet - spin triplet gap and its dependence on both the magnetic field and the confinement potential in the rings.

We restricted ourselves to the two electron system of two-dimensional triple concentric quantum rings where a perpendicular magnetic field is applied, and for two extremes of confinement potential in the rings. We will first present our physical model, and then in the second part of this paper we will study the single electron states in triple concentric quantum rings by using the finite difference method. After that we calculate, using the configuration interaction method, the two electron spin singlet and spin triplet ground and excited state energies, and the spin singlet spin triplet energy gap as a function of the applied magnetic field for both strong and weak confinement. In the last part of our paper, we will discuss the influence of the spin Zeeman energy, by taking different values of the Land\'{e} $g$-factor, on the spin singlet - triplet transition.
\section{\label{sec:2}Model}
In two-dimensional circularly symmetric concentric rings, the single particle Hamiltonian in the presence of a perpendicular magnetic field $B$, using the symmetric gauge $A=-\frac{B}{2}(-y,x,0)$, is given by (in polar coordinates)
\begin{eqnarray}
H_{e}= &-&\frac{\hbar^2}{2m_{e}}\left(\frac{d^2}{d\rho^2}+\frac{1}{\rho}\frac{d}{d\rho}\right)+\frac{\hbar^2l^2}{2m_{e}\rho^2}\nonumber\\
&+&\frac{m_{e}\omega_c^2\rho^2}{8}+\frac{1}{2}\hbar\omega_c l++\mu_B g s_z+V_{e}(\rho)
\label{eq:SHamil},
\end{eqnarray}
where $m_{e}$ is the effective electron band mass, $l$ is the angular momentum of the considered state, and $\omega_c=q_{e}B/m_{e}$ is the cyclotron frequency. Because of circular symmetry the single particle wave function is assumed to be $\Phi_{n,l}=1/\sqrt{2\pi}e^{-il\varphi}\psi_n(\rho) $ where the angular momentum is a conserved quantity. Eq.~(\ref{eq:SHamil}) is the Hamiltonian for the radial part and the problem is thus reduced to an effective one dimensional ($1$D) one. The last term $V_{e}(\rho)$ is the confinement potential of the electron in the concentric rings which is modeled by a parabolic confinement potential and taken of the form:
\begin{equation}
V(\rho)=\frac{m_{e}\omega^2}{2}min[(\rho-R_1)^2,(\rho-R_2)^2,(\rho-R_3)^2,...]
\end{equation}
where $R_i$ stands for the ring radii, $\omega$ is the harmonic oscillator frequency of the lateral confinement for the electrons in each ring\cite{Szafran2005}, and $\mu_B$ is the Bohr magneton. The confinement is taken the same for all rings in order to limit the number of parameters (The outer ring in a real system may have a larger confinement potential, we will discuss its effect at the end of this paper). The second, third, fourth, and fifth terms of the Hamiltonian (\ref{eq:SHamil}) are the centrifugal, diamagnetic, orbital Zeeman terms, and the spin Zeeman term, respectively. The spin-orbit coupling is not considered here as we are dealing with quantum rings of size hundred nanometers. Moreover, the spin Zeeman interaction is decoupled from the orbital degree of freedom, it does not influence the tunnel coupling and can be trivially accounted for as an energy shift linear in $B$. Meanwhile, the Land\'{e} $g$-factor is not a constant in nano-structure (even the sign can changed), but can be controlled through the material concentration\cite{Porras-Montenegro2010,Salis2001}, structure size\cite{Dios-Leyva2006,Babayev2009,Hannak1995}, temperature~\cite{Oestreich1995} and the magnetic field~\cite{Porras-Montenegro2010} applied to the structure. Therefore, in the first part of our paper, we will not take the spin Zeeman energy into account, but the influence of the spin is not neglected and enters through symmetry of the wave function. However, the spin Zeeman energy will have a considerable effect on the spectrum especially for large value of the magnetic field, it lowers~\cite{Wagner1992} (depending on the value of the Land\'{e} $g$-factor) the spin triplet state energy level. Therefore, we will come back to this term and study the influence of it on our results in the last part of the paper.

The single particle Schr\"{o}dinger equation can not be solved analytically. Therefore, we used the finite difference method. The single electron energy is $E_{n,l}$ with wavefunction $\Phi_{n,l}\left(\vec{r}\right)$ where the quantum number $n$ is for the radial part and $l$ is the angular momentum. After obtaining the single electron eigen-states, we can construct the two electron wave functions as a linear combination of products of single electron states $\Psi\left(\vec{r}_1,\vec{r}_2\right)=\Sigma_iC_i\Phi_{n_{1i},l_{1i}}\left(\vec{r}_1\right)\Phi_{n_{2i},l_{2i}}\left(\vec{r}_2\right)$. Then by using the configuration interaction method we can find the eigen-states for two electrons whose Hamiltonian is given by:
\begin{equation}
H_t=H_1+H_2+V_c\left(\vec{r}_1,\vec{r}_2\right)
\end{equation}
where $V_c\left(\vec{r}_1,\vec{r}_2\right)=q_1q_2/4\pi\epsilon\epsilon_0|\vec{r}_1-\vec{r}_2|$ is the Coulomb
potential. We adopted the same method we used previously~\cite{Li2011} to transform the 4-D Coulomb integral into a 2-D one. Note that for the two electron singlet state we should have $\Psi\left(\vec{r}_1,\vec{r}_2\right)=-\Psi\left(\vec{r}_2,\vec{r}_1\right)$, while for the two electron triplet case we have $\Psi\left(\vec{r}_1,\vec{r}_2\right)=\Psi\left(\vec{r}_2,\vec{r}_1\right)$.

In our numerical calculation, we use the GaAs value for the effective mass $m_e=0.067m_0$, the dielectric constant $\epsilon=12.4$, and assume the lateral confinement to be $\hbar\omega=3$ meV, for strong coupling between the different rings and $\hbar\omega=30$ meV for weak coupling for fixed radial positions $R_i$ of the rings.
\section{\label{sec:3}Single particle states}
\begin{figure}[h]
\includegraphics[width=8cm]{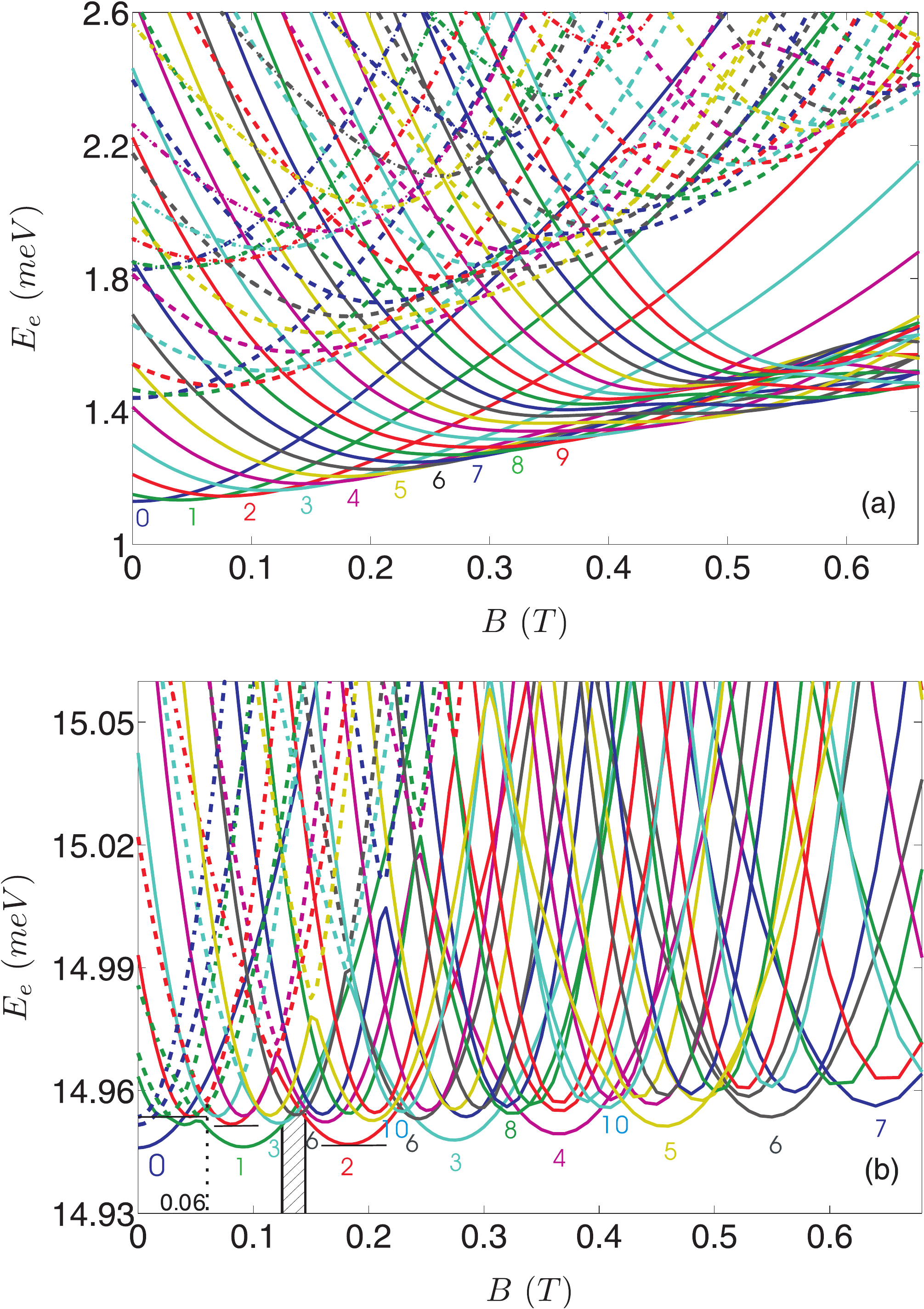}
\vspace{-0.1cm}\caption{\label{fig:Eee3r}(Color online) Single electron spectrum for triple concentric rings with $\hbar\omega=3$ meV (a) and $\hbar\omega=30$ meV (b), as a function of magnetic field. Here solid curves are for the ground state energy of each angular momentum, while dashed and dot-dashed curves are for the first and second excited states, respectively.}
\end{figure}
The results of the single electron eigen-energies for triple concentric rings are shown in Figs.~\ref{fig:Eee3r}(a) (the weak confinement case $\hbar\omega=3$ meV) and \ref{fig:Eee3r}(b) ($\hbar\omega=30$ meV). For the ring radius we took typical values that are realizable experimentally: $R_1=120$ nm, $R_2=180$ nm and $R_3=240$ nm. In Figs.~\ref{fig:Eee3r}(a) and \ref{fig:Eee3r}(b), the solid lines correspond to the ground states of each angular momentum $l$, while the dashed and dot-dashed lines are the corresponding first and second excited states, respectively. It is remarkable that the spectrum of the electron shows a quite different behavior for different confinement potential. The spectrum for $\hbar\omega=30$ meV is similar to a combination of the spectrum of three 1D rings (i.e. rings with zero width). While for $\hbar\omega=3$ meV, it exhibits clear finite width effects (ground state energy far lower than $1/2\hbar\omega$). Comparing with the results of Ref.~\onlinecite{Szafran2005} for CDQRS, we notice that the spectrum shown in Fig.~\ref{fig:Eee3r}(a) is similar to Fig. 3(c) of Ref.~\onlinecite{Szafran2005}, which implies a strong coupling between different rings, while the latter one looks rather like Fig. 3(a) of Ref.~\onlinecite{Szafran2005}, except with a smaller period of the level crossings. This difference results from the different confinement potential which makes the width of the considered rings for the previous case $d_1=2\sqrt{2\hbar/m_e\omega}=54.9$ nm  much larger (for the latter case we have $d_2=17.4$ nm). As a consequence the coupling between the rings is much stronger for the weak confinement case. This is the reason why the electron ground state energy is much lower than $1/2\hbar\omega$.

\begin{figure}
\includegraphics[width=7.8cm]{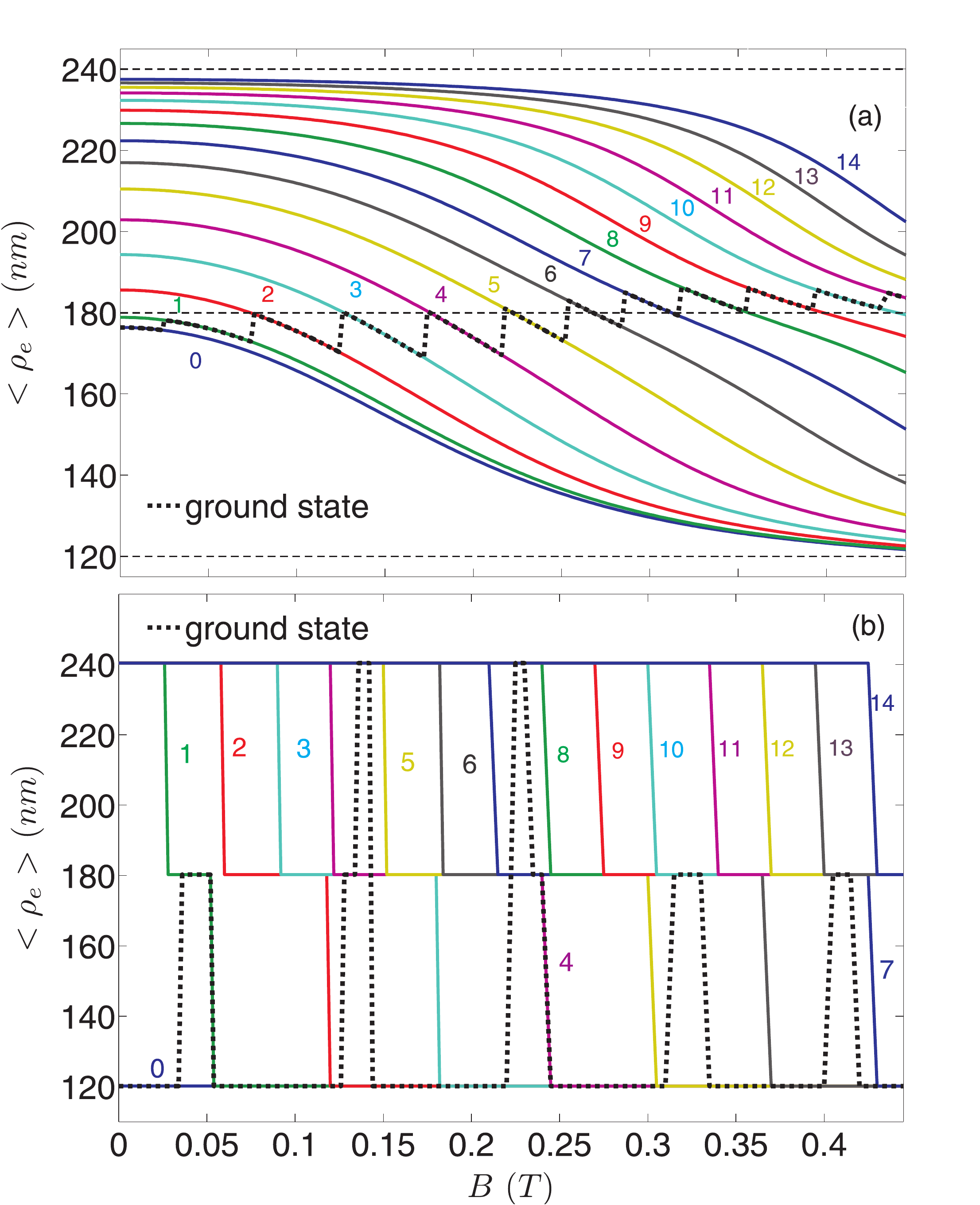} 
\vspace{-0.2cm}\caption{\label{fig:effrho3r}(Color online) Effective radius $<\rho_e>$ of the electron for $\hbar\omega=3$ meV (a) and $\hbar\omega=30$ meV (b) in the lowest energy states of the different angular momentum. The thin dashed lines in (a) shows the position of the rings. The bold dashed dark curves in both figures (a) and (b) show the effective radius of the ground state, while the numbers specify the angular momentum.}
\end{figure}

In both cases, the energy levels of the same angular momentum switch their order through avoided crossings at different magnetic field. And with increasing magnetic field, the energy levels for different angular moment $l$ (except some states with small value of $l$ in the case $\hbar\omega=3$ meV) possess three minima (e.g., see the short horizontal line in Fig.~\ref{fig:Eee3r}(b) for the $l=2$ state), before and after the two avoided level crossings. At those magnetic fields (where levels cross) the corresponding wave functions change. The spatial location of the wave functions change from the outer ring (region) to the inner ring with increasing magnetic field, but this behavior for the two cases is different. This difference is clearly seen from the effective radius of the electron $<\rho_e>$ as shown in Fig.~\ref{fig:effrho3r}. We notice that for the weak confinement case, as the coupling between the rings is strong, the ground states with small value of $l$ (like $l=0,1,2$, which have small centrifugal potentials) exhibit only one clear minimum. As a result, the wave functions locate predominately in the region between the middle and the inner ring at small magnetic field. For states with larger angular momentum $l$, larger centrifugal potentials result in an electron localization that is mainly in the middle ($l=3$, two minima for the energy), or between the outer and the middle ring ($l>3$ and $<6$, two clear minima), or in the outer ring ($l>6$, three minima). For states with larger angular momentum, a larger magnetic field is needed to make the electron wave function leave the outer ring and to transfer entirely to the inner ring, which is clearly shown in Fig.~\ref{fig:effrho3r}(a).
\begin{figure}
\includegraphics[width=8cm]{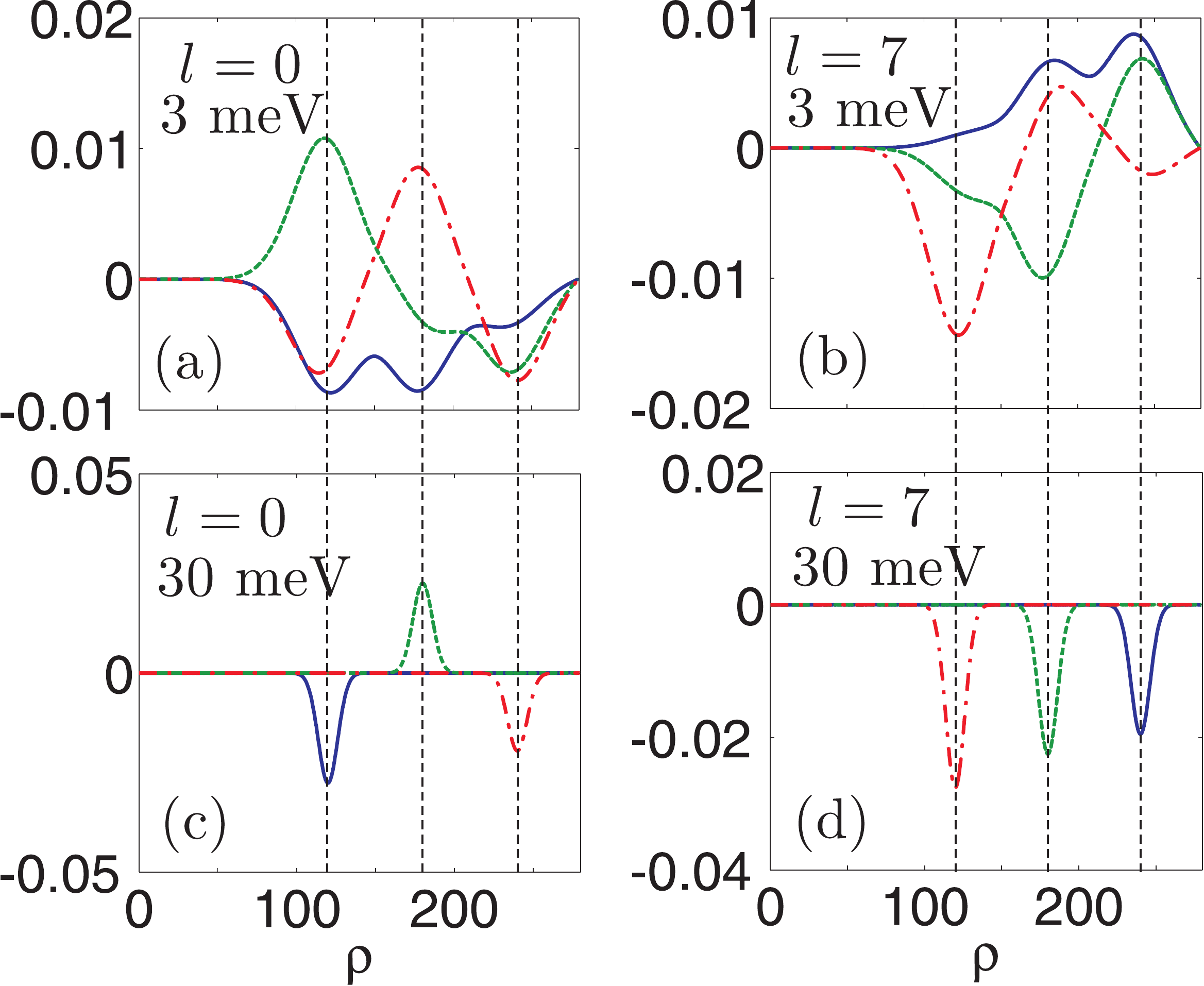} 
\vspace{-0.2cm}\caption{\label{fig:singwave}(Color online) The wave function of the ground (blue solid), first excited (green dashed) and second excited (red dot-dashed) electron states for both weak (upper two figures) and strong confinement (lower two figures) cases for magnetic field $B=0.1$ T, and different angular momentum. The vertical dashed lines indicate the position of the rings (i.e. the minima of the confinement potential).}
\end{figure}

For case $\hbar\omega=30$ meV, the magnetic field dependence of the spectrum is quite different. Small effective width of the rings makes the confinement extremely large and the coupling between the rings very small. As a result the lowest three states for each angular momentum are entirely localized in the three different rings. Fig.~\ref{fig:singwave} shows the lowest three states of the electron wave functions for both the strong and weak confinement cases. Notice that the wave function in the strong confinement case is much more localized, in contrast to the weak confinement case where the wave functions are more extended. We also find from Fig.~\ref{fig:effrho3r}(b) that the $l=0$ state is always localized in the inner ring. While for the non-zero angular momentum states, the electron is entirely in the outer ring (smallest centrifugal potential) for small magnetic field. After reaching the magnetic field where the first avoided level crossing takes place (e.g., $B=0.06$ T for $l=2$ as shown in Fig.~\ref{fig:Eee3r}(b)), the electron switches to the middle ring. And finally, the electron will localize in the inner ring. This can be clearly seen in Fig.~\ref{fig:effrho3r}(b) where $<\rho_e>$ exhibits a steep change after the two avoided level crossings.

There are several additional differences between Figs.~\ref{fig:Eee3r}(a) and \ref{fig:Eee3r}(b): 1) As a result of the extremely small coupling and large ring radii, the distance between the anti-crossing energy levels in the latter case is much smaller (almost zero). 2) For the strong confinement case, there are more frequent angular momentum transitions slightly above the ground state transition, which correspond to the situation where the electron is entirely in the outer or middle ring. Moreover, parts of these transitions can become the ground state transition at some magnetic field (e.g. black dotted region in Fig.~\ref{fig:Eee3r}(b)). 3) Different from the weak confinement case, the angular momentum transition does not exhibit uniform behaviour, it is not always from the electron localized in the same ring and not always with increasing angular momentum. For example, Fig.~\ref{fig:effrho3r}(b) clearly shows that the angular momentum changes in the sequence of ($0$, inner ring) (electron has a zero angular momentum and localized mainly in the inner ring), ($1$, middle ring), ($1$, inner ring), ($3$, middle ring), ($6$, outer ring), ($2$, inner ring) and so on. The state ($0$, inner ring) is the ground state at very small magnetic field. With increasing magnetic field, the Zeeman term decreases more the energy for states with larger angular momentum, while the centrifugal term and the diamagnetic term increase the energy. The centrifugal term decreases the energy of the state that is localized in the ring with larger radius, while the diamagnetic term influences it in the reverse way. Thus, there is a competition between the different wave function configurations in the three rings, the sum of the centrifugal, the diamagnetic and the Zeeman terms in Hamiltonian (\ref{eq:SHamil}) decides which one has the smallest energy (Zeeman term results in an angular momentum transition, and the diamagnetic and centrifugal term determines in which ring the electron is localized).

\begin{figure}
\includegraphics[width=8cm]{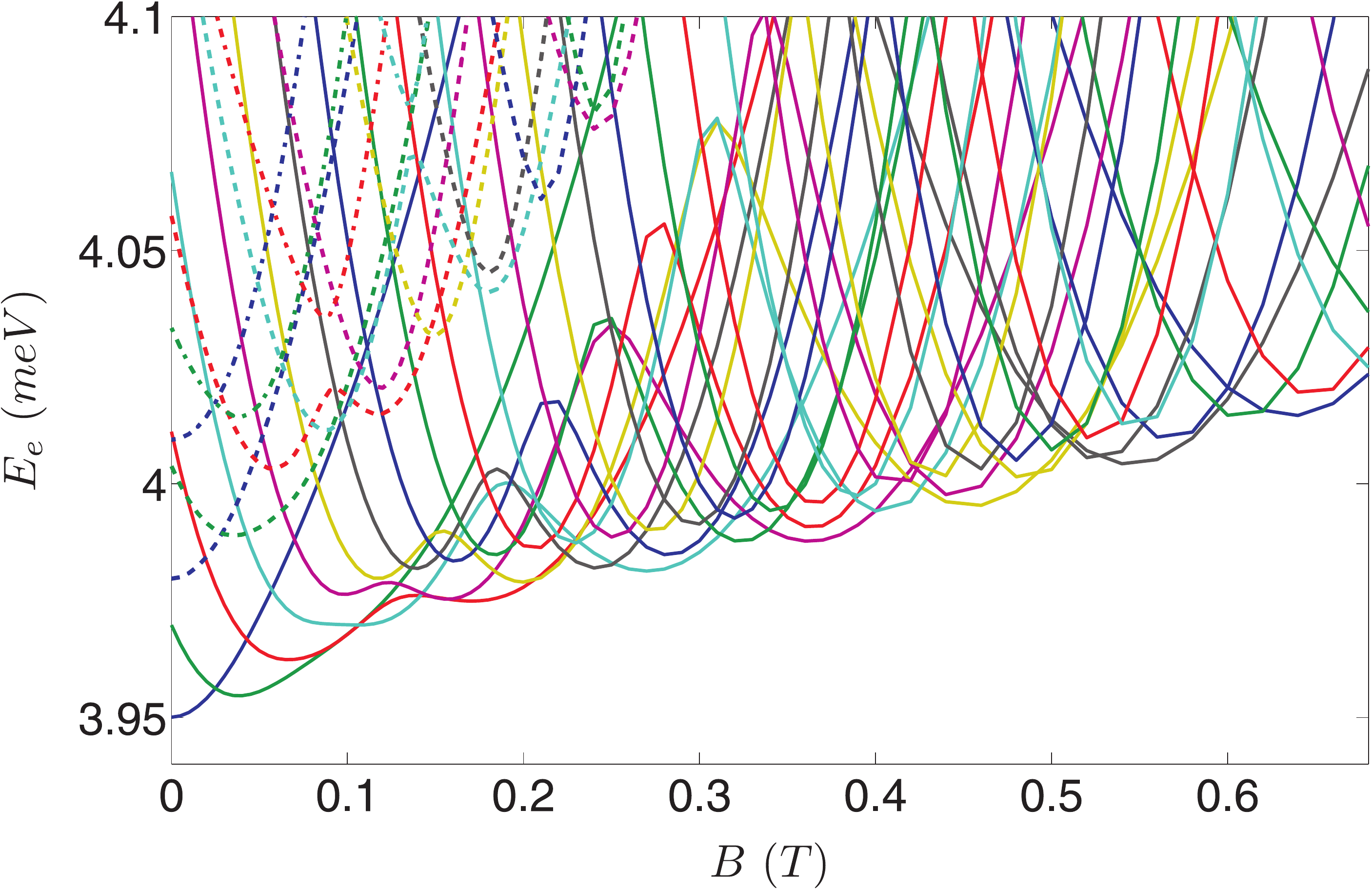}
\vspace{-0.2cm}\caption{\label{fig:Ee83r}(Color online) Single electron spectrum for triple concentric rings with $\hbar\omega=8$ meV, as a function of magnetic field.}
\end{figure}

The above two cases are just two opposite extremes for triple concentric quantum rings. The spectra and effective radii show how the electron behaves at these extreme conditions. We may find an interesting spectrum for the case with medium confinement potential, as shown in Fig~\ref{fig:Ee83r}. The spectrum is just a combination of the two spectra depicted in Figs.~\ref{fig:Eee3r}(a) and (b). For small magnetic field, we have a spectrum like in the weak confinement case, while much more complicated angular momentum transitions occur when the magnetic field is large, just like in the strong confinement case. This also implies that the coupling between the rings becomes smaller with increasing magnetic field.

\section{two electrons confined in triple concentric rings}
We first calculate the two electron energy for different total angular momentum in triple concentric rings with $\hbar\omega=30$ meV. The two electron wave function with fixed total angular momentum $L$ are constructed from linear combinations of the one electron wave functions as follows:
\begin{eqnarray}
\Psi&=&\sum _{n_1=1}^{n_m} \sum_{n_2=1}^{n_m} \sum_{\quad l=-l_m}^{l_m}\!\!\!\!\!\!^\prime C_{n_1n_2}^l\left[\Phi_{n_1,(L+l)/2}(\vec{r_1})\Phi_{n_2,(L-l)/2}(\vec{r_2}) \right.\nonumber\\
&&\left.\pm\Phi_{n_2,(L-l)/2}(\vec{r_1})\Phi_{n_1,(L+l)/2}(\vec{r_2})\right],
\label{eq:phi}
\end{eqnarray}
where the subscript $n_1$ and $n_2$ correspond to the energy levels of the one electron problem with fixed angular momentum. The sum $\sum^\prime$ denotes that only even values of $l$ are taken when $L$ is even and odd values otherwise. Because the spatial and the spin parts of the wave function decouple, the value of the total spin will determine the symmetry of the spatial wave function under particle permutation. The plus sign in Eq.(~\ref{eq:phi}) corresponds to the singlet state with total spin $S=0$, and the minus sign corresponds to the triplet state with $S=1$. Notice that for the spin triplet state, the quantum numbers $n$ and $l$ for the two electrons can not be equal simultaneously. The value $l_m$ is taken sufficiently large such that the energy levels are determined within a given tolerance. As a result of the strong confinement potentials in the rings, the third and fourth excited states have a much larger energy than the lowest three levels. We found that taking $n_m=3$ gives already good accuracy.

\begin{figure}
\includegraphics[width=8cm]{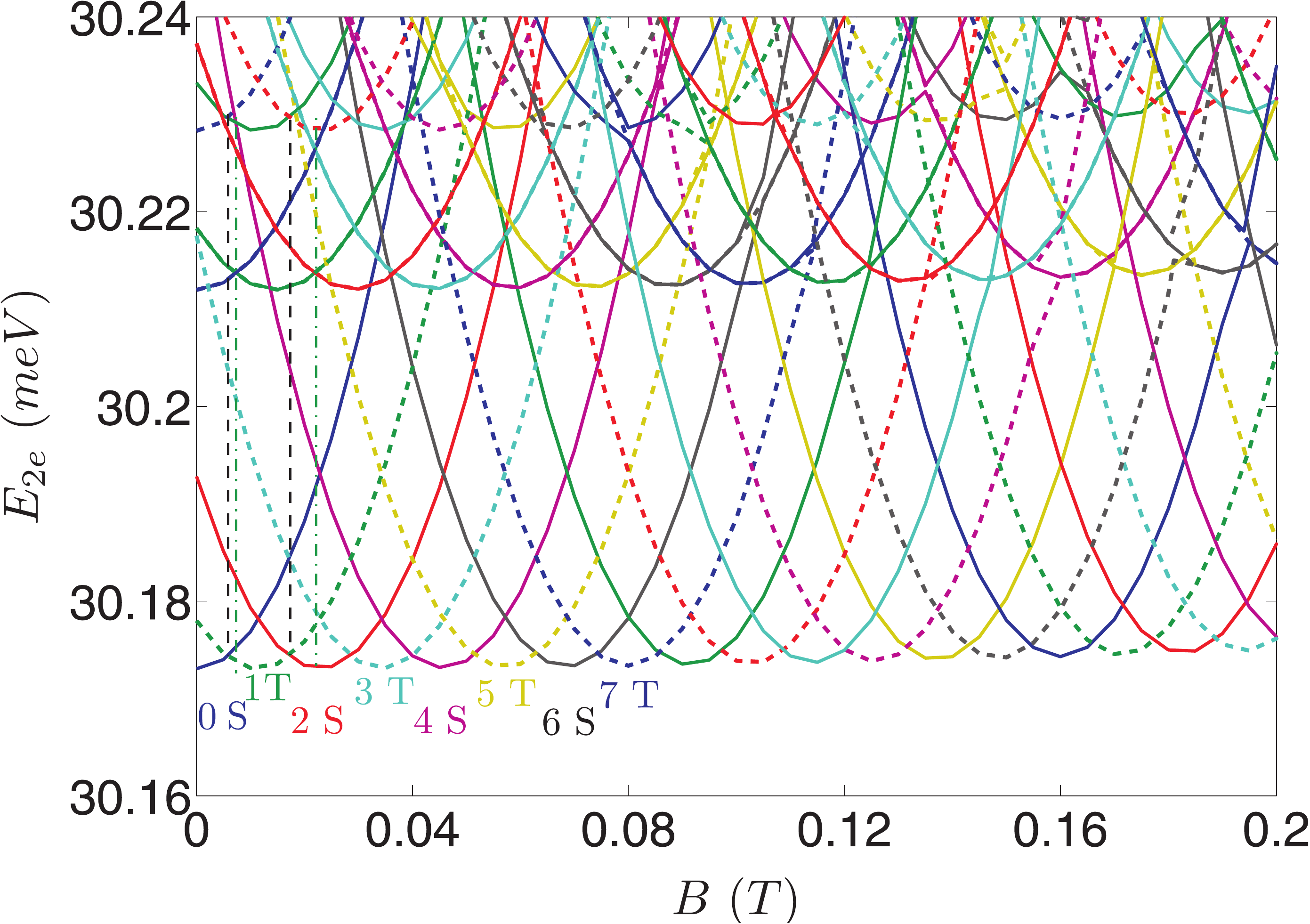}
\vspace{-0.2cm}\caption{\label{fig:E2eL}(Color online) The spin singlet (solid line) and spin triplet (dashed line) states ground state and first exited state energies for different value of total angular momentum $L$, as a function of magnetic field. Different colors correspond to different value of $L$. Here $\hbar\omega=30$ meV.}
\includegraphics[width=8cm]{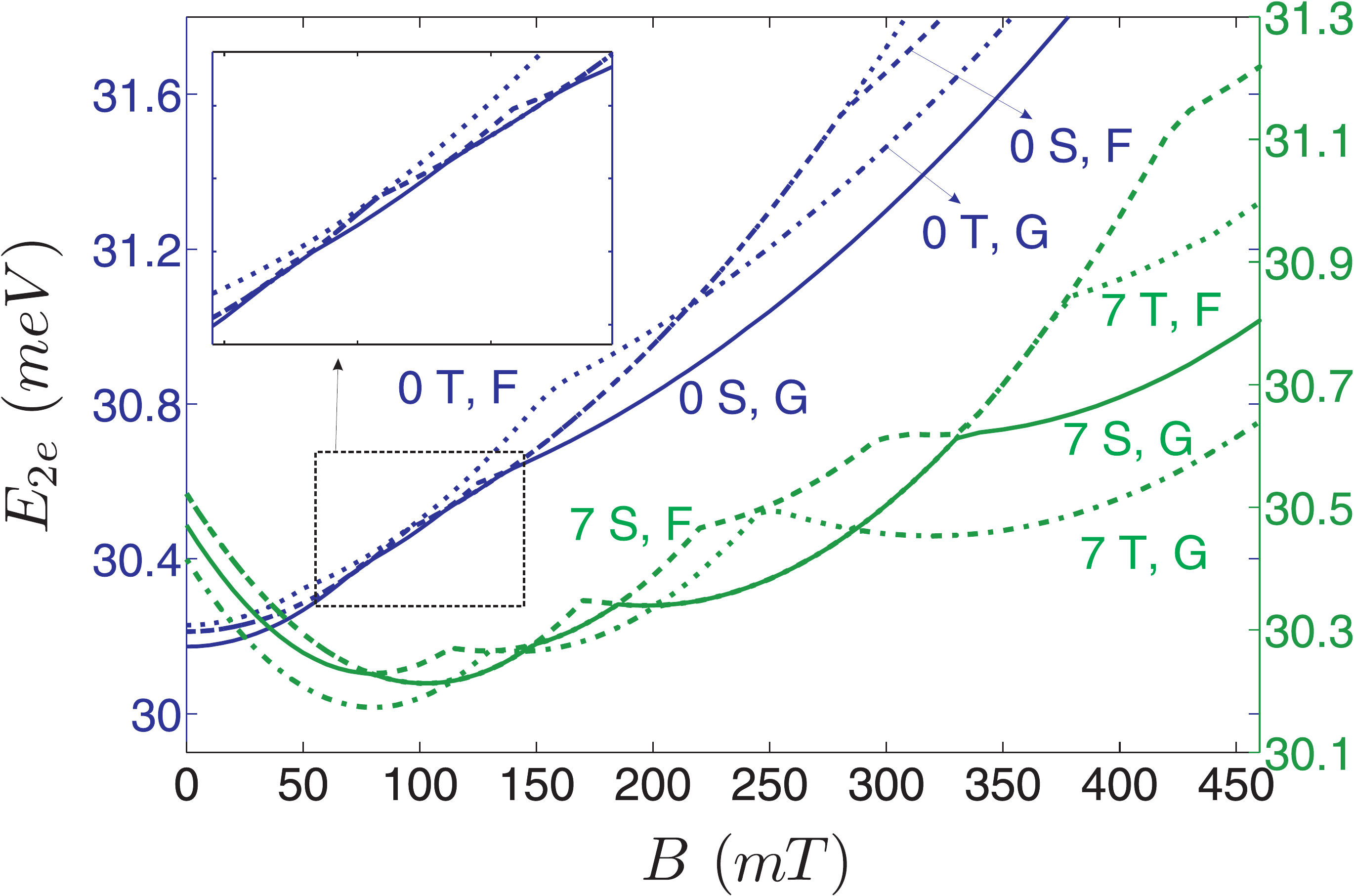}
\vspace{-0.2cm}\caption{\label{fig:EofL07}(Color online) The lowest four states for two electrons in triple concentric rings for $\hbar\omega=30$ meV.  The blue curves are for total angular momentum $L=0$ and green curves for $L=7$. Here S (T) stands for spin singlet (triplet) state, while G (F) stands for the ground and first excited state for the corresponding spin singlet and spin triplet states.}
\end{figure}
The results of the two electron ground and three lowest excited states for different total angular momentum $L$ are shown in Fig.~\ref{fig:E2eL} (always with two singlet states and two triplet states for fixed $L$). The solid (spin singlet) and dashed (spin triplet) line in the same color correspond to the same total angular momentum. Notice that the first and second excited states are degenerate here, however this degenerate behavior only exists in some region of magnetic field (compare Figs.~\ref{fig:E2eL} and \ref{fig:EofL07} for the $L=7$ states). We should mention, as a result of the exchange energy, that for even total angular momentum the spin singlet state is the two electron ground state, while for odd total angular momentum the spin triplet is the ground state. By increasing the magnetic field, the angular momentum of the ground state shifts from $0$ to a larger value and exhibits a fractional Aharonov-Bohm effect. Moreover, the ground state switches between spin singlet and spin triplet states and exhibits a clear spin singlet - spin triplet transition. This transition exists not only in the ground state, but also in the excited states. Notice that from a further inspection of the spectrum we see that the ground state angular momentum transition has the same period (indicated by the vertical dashed line) as the third excited state, but smaller than the overlapped first and second excited states (dotted line). The period for the ground and third excited states is about $B=0.0113$ T, while the period for the first and second excited states is $B=0.0147$ T. Comparing these periods with the single electron spectrum, we find that the period of the ground state $B=0.0113$ T is almost the same as one half the period for the single electron in the most outer ring ($B=0.0224$ T, see the third lowest angular momentum transition in Fig.~\ref{fig:Eee3r}). This indicates that the electrons are in the most outer ring for the two electron ground and third excited states, and is also a fingerprint of the so called fractional Aharonov-Bohm effect. On the other hand, the period of the first and second excited states are much smaller (larger) than one half of the period for the single electron in the middle (outer) ring (which is $0.0407$ T for the middle ring), which means that the electrons are in the area between the middle and outer rings (since the confinement is extremely strong, an electron may be partly in the middle ring and partly in the outer ring).

Figure~\ref{fig:EofL07} shows the lowest four states for two electrons: the spin-singlet ground state (solid curves), the spin-singlet first excited state (dashed line), the spin-triplet ground state (dot-dashed line) and the spin-triplet first excited state (dotted curves). The blue curves (also the ones specified in the inset) are for total angular momentum $L=0$ (even value) and green curves for $L=7$ (odd value). We found that the ground state for $L=0$ is a spin singlet at small magnetic field, and almost overlap with the spin triplet when $B$ reaches $0.074$ T. The inset shows that there are three anti-crossing points between the ground (spin singlet) and the first excited states (spin triplet). The case for $L=7$ is a little different. The spin triplet takes the place of the spin singlet in case of $L=0$, and instead of an anti-crossing, the spin triplet state overlaps with the spin singlet state in some regions of field $B$.

\begin{figure}
\includegraphics[width=8cm]{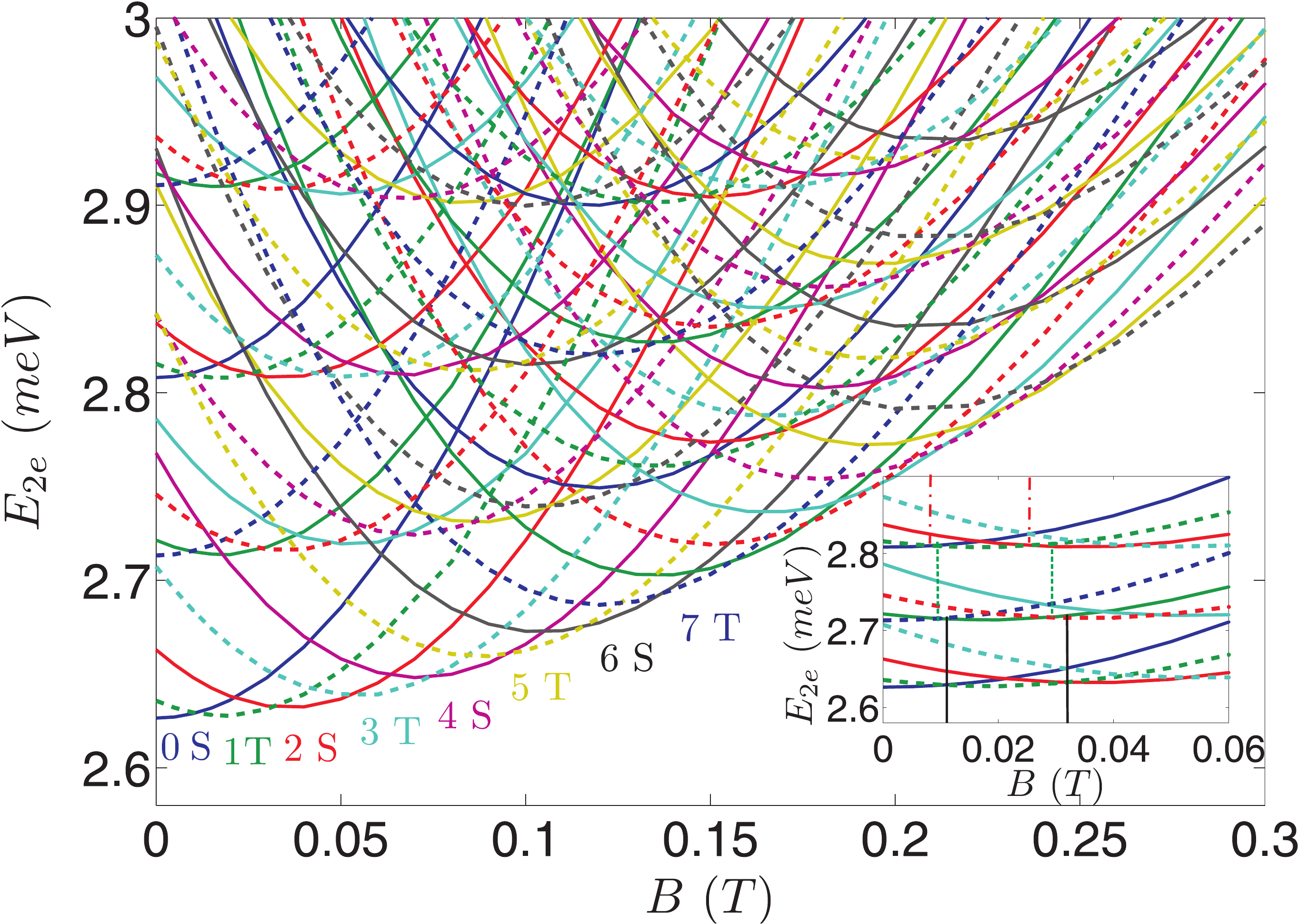}
\vspace{-0.2cm}\caption{\label{fig:E2eL3}(Color online) The spin singlet (solid line) and spin triplet (dashed line) ground state and first exited state energies for the case of $\hbar\omega=3$ meV. In the inset we indicate the period for the lowest three angular momenta transition}
\end{figure}

\begin{figure}
\includegraphics[width=7.5cm]{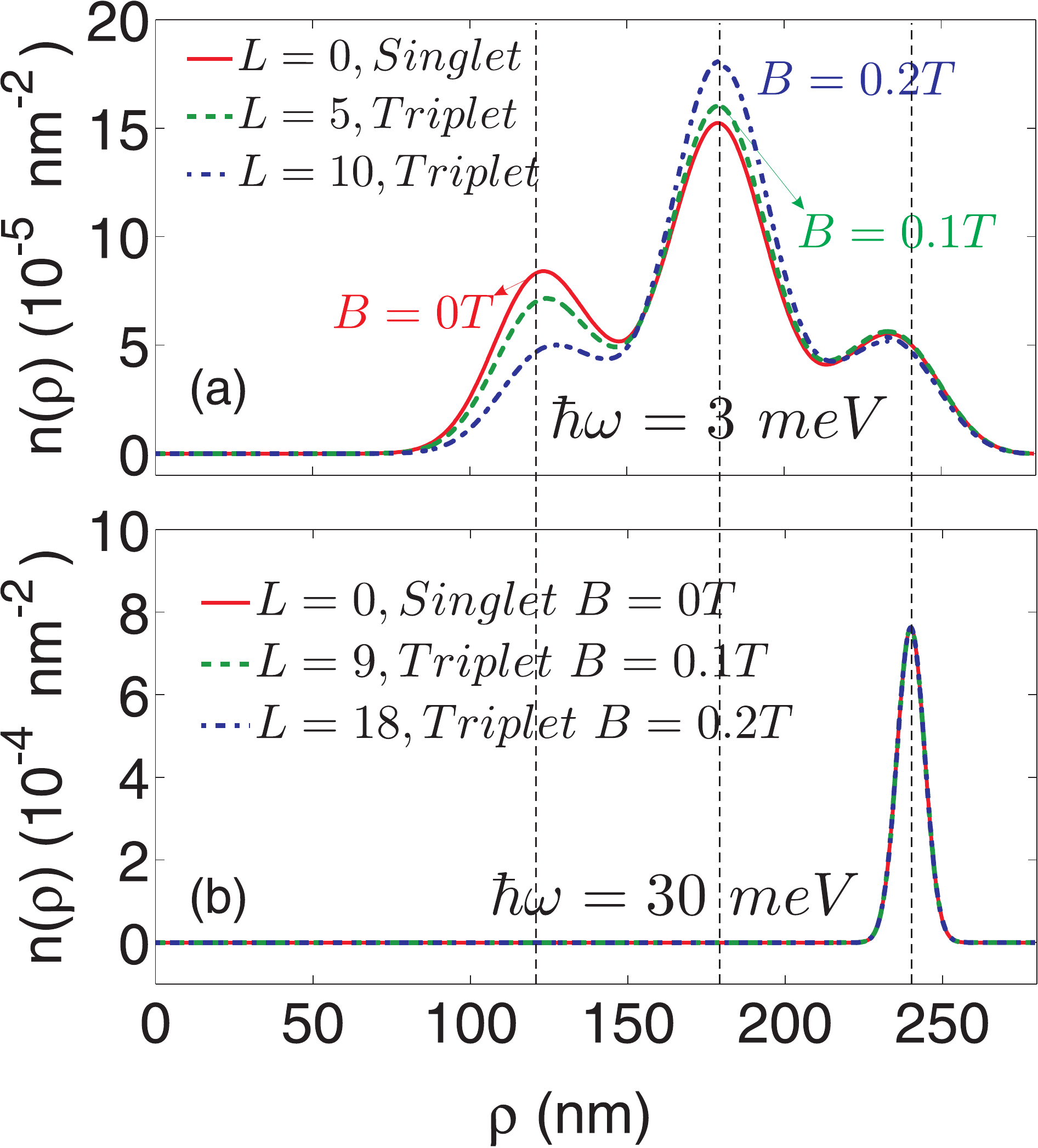}
\vspace{-0.2cm}\caption{\label{fig:denp}(Color online) The electron density in the radial direction of the ground state at three different values of the magnetic field for both (a) $\hbar\omega=3$ meV and (b) $\hbar\omega=30$ meV. The total angular momentum and spin of the corresponding states are also given in the figures. }
\end{figure}

The result for $\hbar\omega=3$ meV is shown in figure~\ref{fig:E2eL3}. As the coupling between the different rings is stronger we take $n_m=6$ instead. Just like the case of strong confinement, the ground state shows a fractional Aharonov-Bohm effect and a similar spin singlet - spin triplet transition. The differences from the previous case are as follows: 1) The second and third angular momentum transition are never degenerate. The degeneracy of the first and second excited states which we have in the strong confinement case, within the given magnetic field region, is lifted; 2) The two electron energy has a distinctly increscent tendency with increasing magnetic field for both ground and excited states. The reason is that for the weak confinement case many more single electron levels have to be taken into account, especially at large magnetic field, and the Coulomb interaction between them strongly increases the total energy. 3) From the inset of Fig.~\ref{fig:E2eL3} we observe that the period for these different transitions are different. The lowest angular momentum transition (ground state) has the largest period, which means that the two electrons occupy the region with the smallest radius as compared to the excited states. Moreover, the period of the ground state transition for $\hbar\omega=30$ meV almost does not change with increasing magnetic field (up to $0.5$ T). While the period for the case of $\hbar\omega=3$ meV has a slight but observable change. For example, the period is $B=0.0211$ T for the transition from $L=1$ to $L=2$, but $B=0.0185$T which is smaller for the transition from $L=13$ to $L=14$ state. This indicates that, in contrast to the case of strong confinement, the two electrons could move slowly to the outer region of the rings with increasing magnetic field.
This can be more clearly seen from the electron density in Fig.~\ref{fig:denp}. The electron probability density in the radial direction is defined by
\begin{equation}
n(\rho)=\sum_{i=1}^2<\delta(\vec{\rho}-\vec{\rho}_i)>.
\end{equation}
We show here the electron probability density of the ground state in the radial direction, for three different values of the magnetic field $B=0$ T (red solid curve), $B=0.1$ T (green dashed curve) and $B=0.2$ T (blue dash-dotted curve). We also give in Fig.~\ref{fig:denp} the value of the angular momentum and total spin of the two electrons ground state at these values of the magnetic field. We find that the electron probability density for strong confinement rings (Fig.~\ref{fig:denp}(b)) overlap for different values of the magnetic field, and it is always concentrated in the most outer ring. While for weak confinement rings, as shown in Fig.~\ref{fig:denp}(a), the probability density peaks of the electron moves slightly to the region with larger $\rho$ with increasing magnetic field. Moreover, the electron probability extend into all the rings as a result of the strong coupling between the different rings. It is predicted that the coupling between the different rings could decrease as a result of the large energy difference in the rings when the magnetic field is very large. Thus we may have a similar spectrum as in the strong confinement case.

\begin{figure}
\includegraphics[width=8cm]{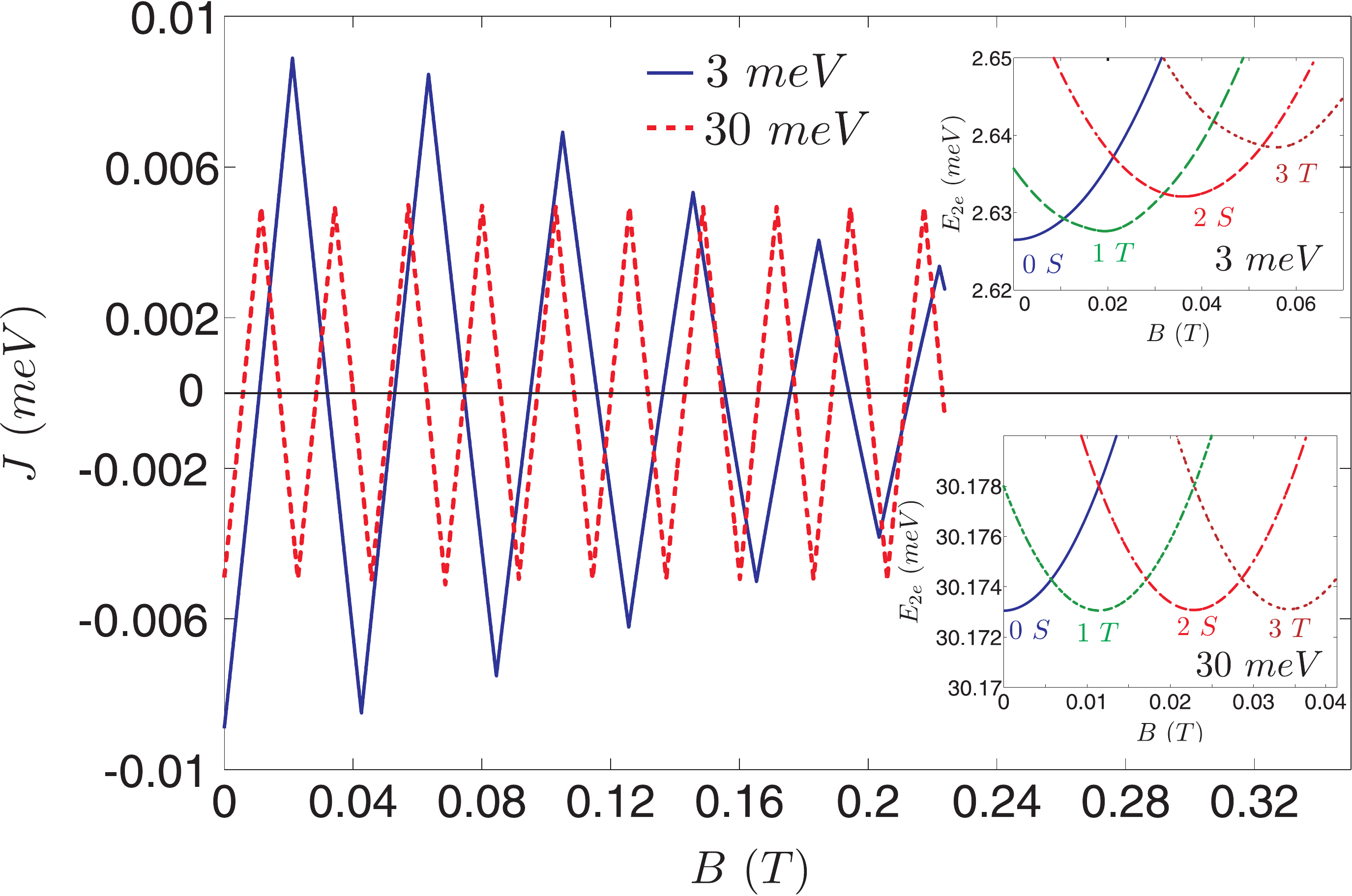}
\vspace{-0.2cm}\caption{\label{fig:E3rsplitE}(Color online) The splitting energy of the spin singlet and spin triplet ground states as a function of the magnetic field. The blue solid line is for $\hbar\omega=3$ meV, while red dashed line for the case of $\hbar\omega=30$ meV. Insets are the ground state transition for these two cases.}
\end{figure}

To have a better comparison of the two cases and to have a better understanding of the spin singlet - spin triplet transition, we show in Fig.~\ref{fig:E3rsplitE}, the splitting energy $J$ (the energy gap of the spin singlet ground state and spin triplet ground state) as a function of the magnetic field. Notice that in both cases the splitting energy is very small (around $0.01$ meV), and oscillates with increasing magnetic field. The splitting energy for $\hbar\omega=3$ meV has a larger amplitude at small value of the magnetic field, but a decreasing amplitude when $B$ increases. The amplitude of the splitting energy for rings with strong confinement is almost constant with increasing magnetic field. As the confinement energy in concentric rings can be changed through an external gate voltage, we could modify both the period of the spin singlet - spin triplet transition and amplitude of the splitting energy by changing the gate voltage. This implies that the singlet - triplet transition can be tuned by both the confinement potential and the magnetic field, and thus it tunes the ground state entanglement of the two electron system. A similar study was done in Ref.~\onlinecite{Kyriakidis2002} for lateral quantum dots with parabolic and non-parabolic confinement, where the authors found a singlet-triplet energy gap that depends on the gate voltage. However the influence of the magnetic field on the singlet-triplet gap amplitude was not discussed. We believe that the decreasing amplitude for the weak confinement rings mainly results from the concentric geometry of the ring where the electrons can extend more to the outer rings by increasing magnetic field. Therefore, similar results may not exist in quantum dots. Our results for TCQRs with strong confinement potential confirm the behavior of the persistent current studied by local-spin density-functional theory (LSDET) in Ref.~\onlinecite{Escartin2010} where TCQRs with only strong confinement were chosen.
\subsection{Influence of Spin Zeeman energy}
\begin{figure}
\includegraphics[width=7.5cm]{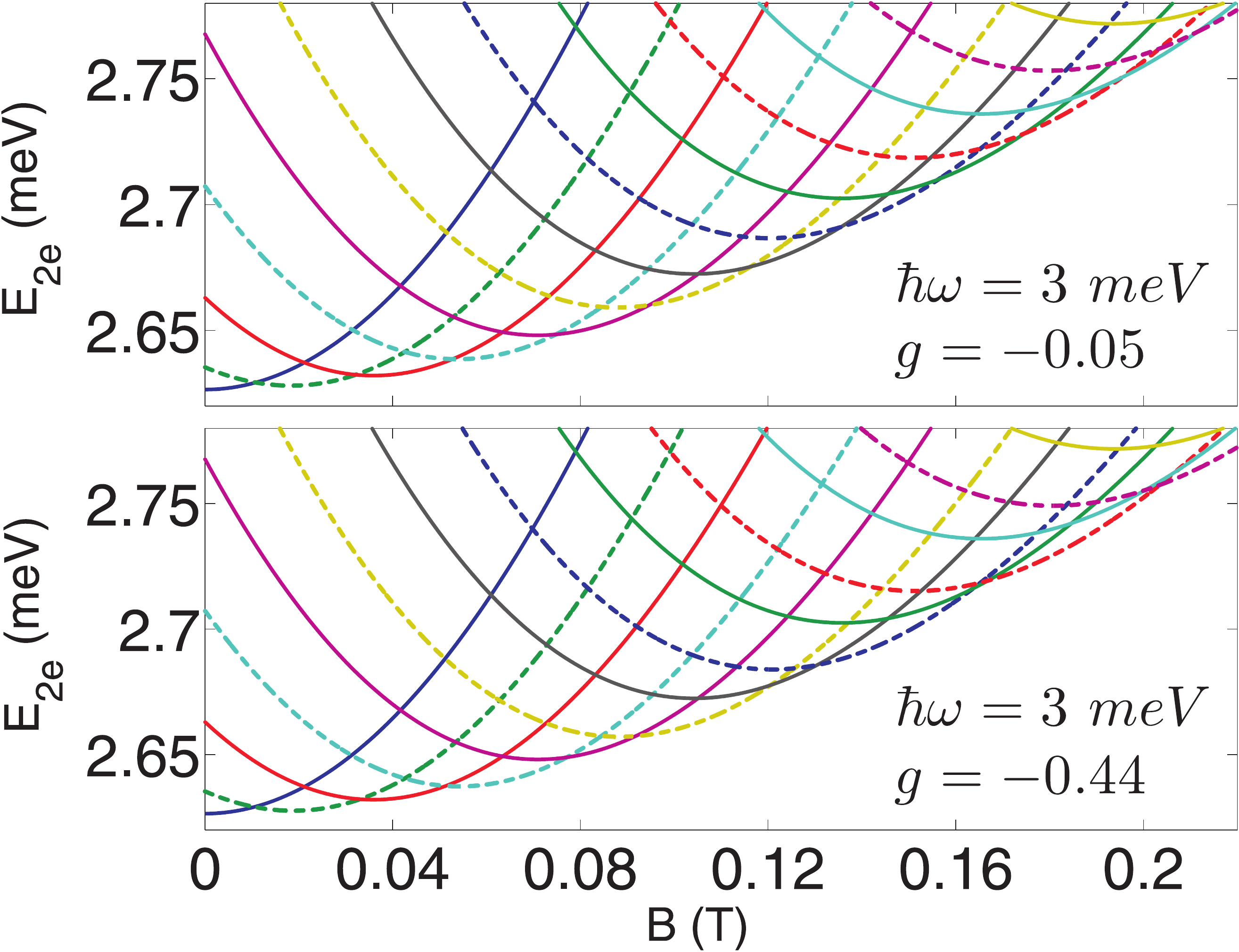}
\vspace{-0.2cm}\caption{\label{fig:E3r3meVg}(Color online) The two electron spin singlet (solid curves) and spin triplet (dashed curves) ground states as a function of the magnetic field, in the presence of the spin Zeeman energy. Here the confinement is $\hbar\omega=3$ meV, and the Land\'{e} $g$-factor is chosen to be $g=-0.05$ (upper figure) and $g=-0.44$ (lower figure).}
\end{figure}

\begin{figure}
\includegraphics[width=8cm]{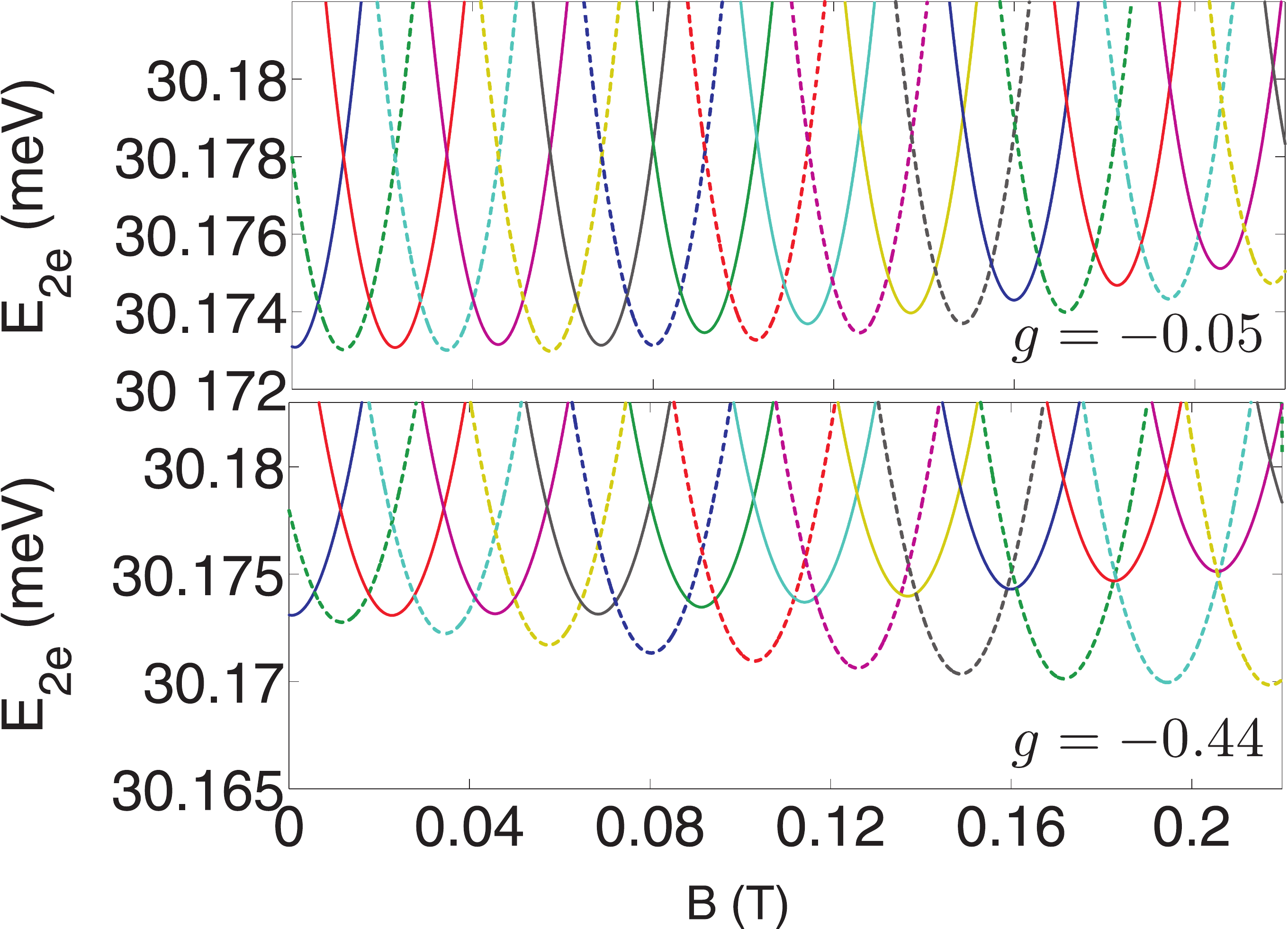}
\vspace{-0.2cm}\caption{\label{fig:E3r30meVg}(Color online) Same as in Figure \ref{fig:E3r3meVg} but now for $\hbar\omega=30$ meV.}
\end{figure}

The triple concentric GaAS/AlGaAs rings we studied can be both fabricated through lithography and self-assembly approaches. In the first case, the magnetic length is typically smaller than the ring radius, and the Land\'{e} $g$-factor is related to those of quantum wells with a value from around $-0.5$ to $0.5$ and depends strongly on the width~\cite{Porras-Montenegro2010,Jeune1997,Hannak1995} of the well and magnetic field~\cite{Porras-Montenegro2010}. For the second case, the Land\'{e} $g$-factor is also not fixed, and was previously taken with completely different values, such as in Refs.~\onlinecite{Kyriakidis2002} and \onlinecite{Escartin2010} where the value $g=-0.44$ of bulk GaAs was taken. But the value of the Land\'{e} $g$-factor is actually strongly material concentration~\cite{Salis2001} and magnetic field~\cite{Porras-Montenegro2010} dependent. Here we assumed an isotropic Land\'{e} $g$-factor with values of $-0.44$ as in bulk GaAs, $-0.05$ and $-0.1$ to calculated the shifted two electron ground state energy, and compare them to the one without the spin Zeeman term. The result is shown in Fig.~\ref{fig:E3r3meVg} for rings with weak confinement potential and in Fig.~\ref{fig:E3r30meVg} for the strong confinement case.

\begin{figure}
\includegraphics[width=8cm]{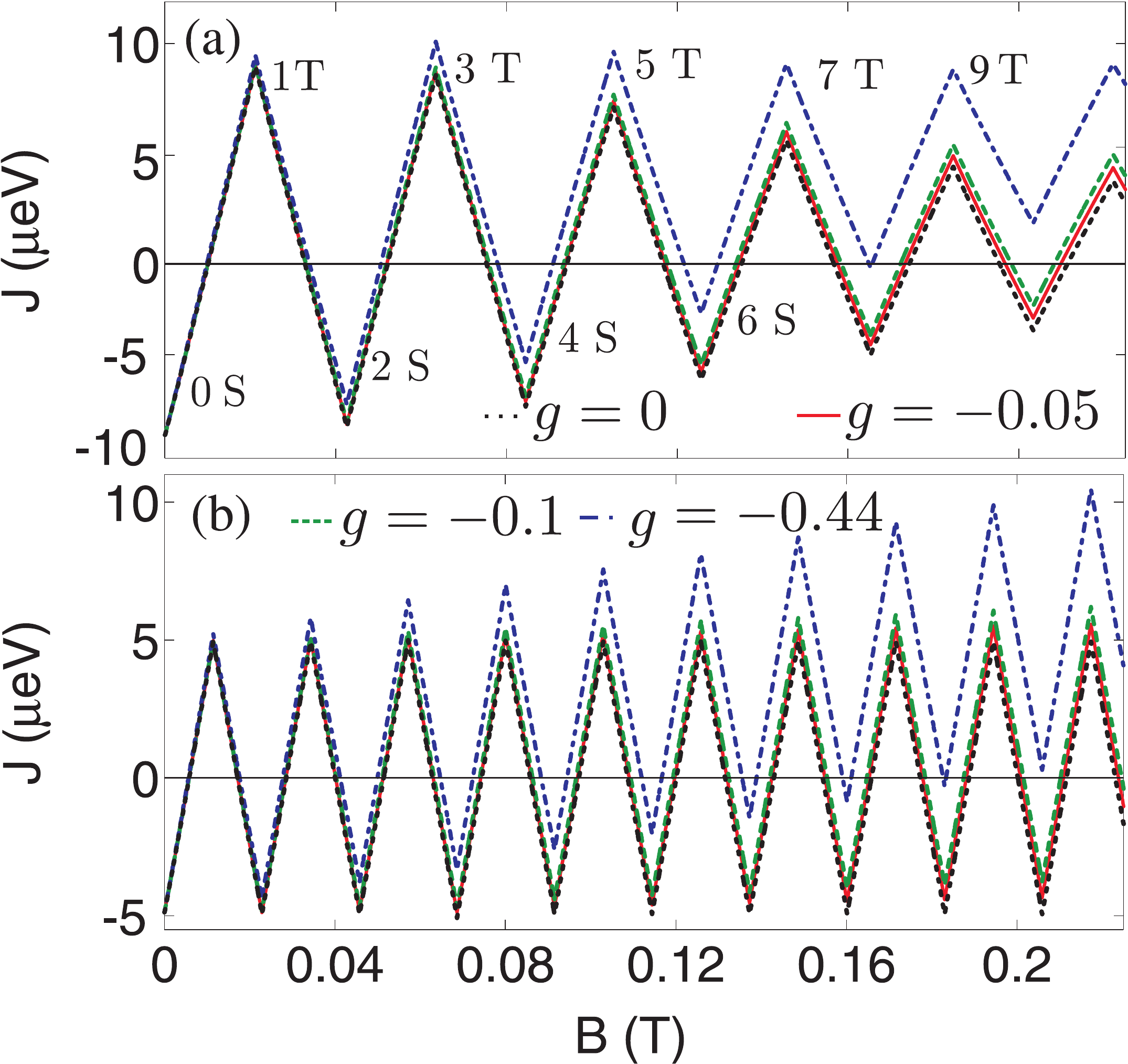}
\vspace{-0.2cm}\caption{\label{fig:Esplit3rg}(Color online) The splitting energy of the spin singlet and spin triplet ground states as a function of the magnetic field, for four different values of the Land\'{e} $g$-factor $g=0$ (black dotted curce), $-0.05$ (red solid curve), $-0.1$ (green dashed curve), and $-0.44$ (blue dot-dashed curve). While figure (a) is for the weak confinement case with $\hbar\omega=3$ meV and figure (b) for the strong confinement $\hbar\omega=30$ meV.}
\end{figure}

In the presence of the spin Zeeman energy, the two electron spin triplet state energy has an additional term $g\mu_BBS_z$. Therefore, the triplet state splits into three distinct levels, here only the lowest triplet state is considered and shown, as it lowers the triplet state energy thus it may change the previous ground state transitions. We found that when we just take the bulk value of $g$ the spin Zeeman energy may largely lower the triplet state energy, especially when the magnetic field is strong. The spin singlet state in this case is strongly suppressed, and even not present in the ground state at large magnetic field (e.g., after $B=0.16$ T for the strong confinement case). When the Land\'{e} $g$-factor is considerable small, for example $g=-0.05$ as shown here, the spin Zeeman energy effect is negligible for a large range of magnetic fields. For a more clear investigation, the splitting energies of the spin singlet and spin triplet ground states in the presence of spin Zeeman energy for different values of Land\'{e} $g$-factor are shown in Fig.~\ref{fig:Esplit3rg}. Here positive $J$ implies that the spin triplet state is the two electron ground state. It shows, beside the case $g=-0.44$, that the spin Zeeman energy does not have a large influence on the splitting energy, especially for small magnetic field. But for the case $g=-0.44$ the effect is considerably large and the triplet state always has the lowest energy at large magnetic field. Thus the total angular momentum does not change continuously, e.g., for the case $\hbar\omega=3$ meV, the total angular momentum of the ground state increase in the sequence $0$, $1$, $2$, $3$, $4$, $5$, $6$, $7$, $9$, $11$ (this is specified in the upper figure together with the corresponding total spin). Notice that as the spin Zeeman energy only introduces a magnetic dependent linear term to the triplet state and this shift is the same for different angular momentum, the spin Zeeman energy will not change the period of the spin singlet (or triplet) angular momentum transition. As a result, the period of the splitting energy is independent of the Land\'{e} $g$-factor, this is clearly shown in Fig.~\ref{fig:Esplit3rg}. But the amplitude decreases linear as a function of magnetic field. Moreover, we found that the spin Zeeman energy has a slightly stronger effect on the weak confinement case.

\section{\label{sec:5}conclusions}
In this paper we studied the coupling between triple concentric rings and its consequences on the two electron spin singlet - spin triplet transition. The effects of the magnetic field and spin-Zeeman energy, by choosing different value of the Land\'{e} $g$-factor, on the splitting energy are also studied. We found that when the confinement potential is weak ($3$ meV) the coupling between the different rings is strong, while the coupling is considerable weaker for the strong confinement case and the spectrum exhibits a richer structure. In contrast to the single ring case, the angular momentum transitions are not always from small to large angular momentum, especially when the confinement is strong. The ground state angular momentum could change from the outer ring with large angular momentum to the inner ring but with small angular momentum, as shown in Fig.~\ref{fig:Eee3r}(b) and Fig.~\ref{fig:Ee83r}. For the strong confinement case, there are angular momentum transitions with different period which correspond to different rings in TCQRs. The sum of Zeeman energy and diamagnetic shift (centrifugal energy is also important, especially for rings with small radii) determines which angular momentum transition becomes the ground state transition. In medium confined rings, we can have an angular momentum transition which is just the combination of the two extremes.

The ground state of the two electrons exhibits an spin singlet - spin triplet transition, and this transition is strongly depended on the confinement potential of the rings (or the related coupling between different rings). By decreasing the confinement energy, we can increase the period of the transition and the amplitude of the anti-crossing energy gap. We also found that this anti-crossing energy gap (splitting energy) for the strong confinement case has an almost constant amplitude (at least within the magnetic field range we have considered). But it has a clear decreasing tendency with increasing magnetic field for rings with weak confinement. As a result, by choosing multiple concentric rings with different confinement potential, we can have distinct ground state entanglements of the two electron system. And this ground state entanglement can be tuned by a magnetic field. This magnetic field and confinement tunable singlet-triplet transition is of paramount importance in previous proposals for quantum-gate operation, because it allows for a controllable spin swap. This behavior can be observed by detecting the persistent current or from a far-infrared study. Similar results for the strong confinement case were previously presented by Escart\'{\i}n et al.~\cite{Escartin2010}, who reported few electron ground states in TCQRs by using local-spin density-functional theory. They restricted themselves to rings with strong confinement and showed a similar magnetic field dependent behavior in the persistent current. Notice that this spin singlet - triplet transition mainly depends on the confinement or the coupling between the rings. The size of rings is also a critical parameter but it rather determines the period of the fractional AB effect. We may find in Ref.~\onlinecite{Szafran2005} for CDQRs with weak confinement that the two electron spectrum has a similar pattern as in TCQRs, although they used different ring radii. The differences of TCQRs from the CDQRs are: 1) The outer ring has a considerable larger radius in TCQRs, thus a smaller magnetic field is required to realize the spin singlet - spin triplet transition. 2) As a result of the increased number of rings, the two electron ground and excited states have more optional configurations with quite different related singlet - triplet periods, and the difference between the ground and first excited state energy should be more sensitive to the magnetic field.

The spin Zeeman energy is also taken into account by choosing a different value of the Land\'{e} $g$-factor. We found that the spin Zeeman energy has a considerable large influence on the magnetic dependent spin singlet - spin triplet transition, especially when the Land\'{e} $g$-factor is large. However, the influence of the spin Zeeman energy is small enough to be neglect when the magnetic field is very small, even for a large Land\'{e} $g$-factor. By increasing the applied magnetic field, the spin triplet state lowers more its energy that the spin singlet state which is now suppressed as the ground state, and finally the triplet state persists to be the ground state. We should mention here that the period of the splitting energy $J$ was independent of the choice of the Land\'{e} $g$-factor, while the amplitude decreased linearly as a function of the magnetic field. Moreover, the TCQR with weak confinement was affected more by the spin Zeeman energy.

The outer ring can exhibit a larger confinement than the inner ring as one expects to be the case in case of self-assembled concentric rings. Such a large confinement may strongly repel the electrons from the most outer ring. In such a case, two main features appear. If the confinement is not very strong such that the electrons can still be localized in the most outer ring, the two electron spectrum possess a similar structure as for a normal TCQRs, but the period of the ground state transitions will be larger. And as the wave function is repelled towards the region between the middle and outer rings, the coupling between these two rings will be stronger, thus each splitting (i.e., ant-crossings) as shown in Fig.~\ref{fig:E3rsplitE} will be just like in the weaker confinement case. But if the confinement in the outer ring is so strong that the electrons can not localize in this ring, the TCQRs will behave like a normal concentric double quantum ring with a less rich structure of the energy spectrum.

\begin{acknowledgments}
This work was supported by the EU-NoE: SANDiE, and the Flemish Science Foundation (FWO-Vl).
\end{acknowledgments}

%

\begin{references}
\bibitem{Loss1998}
Daniel Loss and David P. DiVincenzo.
\newblock Quantum computation with quantum dots.
\newblock {\em Phys. Rev. A}, 57:120--126, Jan 1998.

\bibitem{Barenco1995}
A. Barenco, D. Deutsch, A. Ekert, and R. Jozsa.
\newblock Conditional Quantum Dynamics and Logic Gates.
\newblock {\em Phys. Rev. Lett.}, 74:4083--4086, May 1995.

\bibitem{Brum1997}
J. A. Brum and P. Hawrylak.
\newblock Coupled quantum dots as quantum exclusive-OR gate.
\newblock {\em Superlattices and Microstructures}, 22(3):431 -- 436, 1997.

\bibitem{Kane1998}
B. E. Kane.
\newblock A silicon-based nuclear spin quantum computer.
\newblock {\em Nature}, 393(6681):133, 1998.

\bibitem{Burkard1999}
Guido Burkard, Daniel Loss, and David~P. DiVincenzo.
\newblock Coupled quantum dots as quantum gates.
\newblock {\em Phys. Rev. B}, 59:2070--2078, Jan 1999.

\bibitem{Kyriakidis2002}
Jordan Kyriakidis, M. Pioro-Ladriere, M. Ciorga, A. S. Sachrajda, and
  P. Hawrylak.
\newblock Voltage tunable singlet-triplet transition in lateral quantum dots.
\newblock {\em Phys. Rev. B}, 66:035320, Jul 2002.

\bibitem{Lee2004}
S. D. Lee, S. J. Kim, J. S. Kang, Y. B. Cho, J. B. Choi, Sooa Park, S. R. Eric
  Yang, S. J. Lee, and T. H. Zyung.
\newblock Spin singlet-triplet transition in a Si-based two-electron double
  quantum dot molecule.
\newblock 2004.

\bibitem{Fuhrer2001}
A. Fuhrer, S. Luscher, T. Ihn, T. Heinzel, K. Ensslin, W. Wegscheider, and
  M. Bichler.
\newblock Energy spectra of quantum rings.
\newblock {\em Nature}, 413(6858):822, 2001.

\bibitem{Lorke2000}
Axel Lorke, R. Johannes Luyken, Alexander O. Govorov, J\"{o}rg P. Kotthaus, J. M.
  Garcia, and P. M. Petroff.
\newblock Spectroscopy of Nanoscopic Semiconductor Rings.
\newblock {\em Phys. Rev. Lett.}, 84:2223--2226, Mar 2000.

\bibitem{Szafran2005}
B. Szafran and F. M. Peeters.
\newblock Few-electron eigenstates of concentric double quantum rings.
\newblock {\em Phys. Rev. B}, 72:155316, Oct 2005.

\bibitem{Chakraborty1994}
T. Chakraborty and P. Pietil\"ainen.
\newblock Electron-electron interaction and the persistent current in a quantum
  ring.
\newblock {\em Phys. Rev. B}, 50:8460--8468, Sep 1994.

\bibitem{Niemela1996}
P. Hyvonen. K. Niemiel\"{a}, P. Pietilainen and T. Chakraborty.
\newblock Fractional oscillations of electronic states in a quantum ring.
\newblock {\em Europhys. Lett}, 36:533, Oct 1996.

\bibitem{Wiel2003a}
W. G. van der Wiel, Yu. V. Nazarov, S. De Franceschi, T. Fujisawa, J. M.
  Elzerman, E. W. G. M. Huizeling, S. Tarucha, and L. P. Kouwenhoven.
\newblock Electromagnetic Aharonov-Bohm effect in a two-dimensional electron
  gas ring.
\newblock {\em Phys. Rev. B}, 67:033307, Jan 2003.

\bibitem{Garcia1997}
J.M. Garcia and G. Medeiros-Ribeiro.
\newblock Intermixing and shape changes during the formation of InAs
  self-assembled quantum dots.
\newblock {\em Applied Physics Letters}, 71(14):2014, 1997.

\bibitem{Gong2005}
Z. Gong, Z. C. Niu, S. S. Huang, Z. D. Fang, B. Q. Sun, and J. B. Xia.
\newblock Formation of GaAs/AlGaAs and InGaAs/GaAs nanorings by droplet
  molecular-beam epitaxy.
\newblock {\em Applied Physics Letters}, 87(9):093116, 2005.

\bibitem{Mano2005}
T. Mano, T. Kuroda, S. Sanguinetti, T. Ochiai, T. Tateno, J. Kim, T. Noda, M. Kawabe,
  K. Sakoda, G. Kido, and N. Koguchi.
\newblock Self-assembly of concentric quantum double rings.
\newblock {\em Nano Letters}, {5}(3):425--428, MAR 2005.

\bibitem{Kuroda2005}
T. Kuroda, T. Mano, T. Ochiai, S. Sanguinetti, K. Sakoda, G. Kido, and
  N.~Koguchi.
\newblock Optical transitions in quantum ring complexes.
\newblock {\em Phys. Rev. B}, 72:205301, Nov 2005.

\bibitem{Somaschini2009}
C. Somaschini, S. Bietti, N. Koguchi, and S. Sanguinetti.
\newblock Fabrication of Multiple Concentric Nanoring Structures.
\newblock {\em Nano Letters}, 9(10):3419--3424, OCT 2009.

\bibitem{Somaschini2010}
C. Somaschini, S. Bietti, A. Fedorov, N. Koguchi, and S. Sanguinetti.
\newblock Concentric Multiple Rings by Droplet Epitaxy: Fabrication and Study
  of the Morphological Anisotropy.
\newblock {\em Nanoscale Research Letters}, 5:1865--1867, 2010.
\newblock 10.1007/s11671-010-9699-6.

\bibitem{Climente2006}
J. I. Climente, J. Planelles, M. Barranco, F. Malet, and M. Pi.
\newblock Electronic structure of few-electron concentric double quantum rings.
\newblock {\em Phys. Rev. B}, 73:235327, Jun 2006.

\bibitem{Planelles2005}
J. Planelles and J. I. Climente.
\newblock Semiconductor concentric double rings in a magnetic field.
\newblock {\em European Physical Journal B -- Condensed Matter}, 48(1):65 - 70, 2005.

\bibitem{Escartin2009}
J. M. Escart\'{\i}n, F. Malet, A. Emperador, and Mart\'{\i} Pi.
\newblock Isomeric electronic states in concentric quantum rings.
\newblock {\em Phys. Rev. B}, 79:245317, Jun 2009.

\bibitem{Escartin2010}
M. Barranco J. M. Escart\'{\i}n and Mart\'{\i} Pi.
\newblock Ground state and infrared response of triple concentric quantum ring
  structures.
\newblock {\em Phys. Rev. B}, 82:195427, Nov 2010.

\bibitem{Wagner1992}
M. Wagner, U. Merkt, and A. V. Chaplik.
\newblock Spin-singlet\char21{}spin-triplet oscillations in quantum dots.
\newblock {\em Phys. Rev. B}, 45:1951--1954, Jan 1992.

\bibitem{Porras-Montenegro2010}
N. Porras-Montenegro, J.~Dar\'{\i}o Perea, and J. R. Mej\'{\i}a-Salazar.
\newblock The electron Land\'{e} g factor in GaAs-(Ga, Al)As coupled quantum
  wells.
\newblock {\em AIP Conference Proceedings}, 1199(1):253--254, 2010.

\bibitem{Salis2001}
G.~Salis, Y. Kato, K. Ensslin, D.C. Driscoll, A.C. Gossard, and D.D. Awschalom.
\newblock Electrical control of spin coherence in semiconductor nanostructures.
\newblock {\em Nature}, 414(6864):619, 2001.

\bibitem{Dios-Leyva2006}
M. de Dios-Leyva, N. Porras-Montenegro, H. S. Brandi, and L. E. Oliveira.
\newblock Cyclotron effective mass and Land\'{e} g factor in GaAs Ga1-xAlxAs
  quantum wells under growth-direction applied magnetic fields.
\newblock {\em Journal of Applied Physics}, 99(10):104303, 2006.

\bibitem{Babayev2009}
A.M. Babayev, \"{O}. Mercan, and S. Tez.
\newblock Electron Lande g-factor in GaAs/AlxGa1-xAs quantum wires.
\newblock {\em Physica E: Low-dimensional Systems and Nanostructures},
  41(3):345 -- 348, 2009.

\bibitem{Hannak1995}
R.M. Hannak, M. Oestreich, A.P. Heberle, W.W. R\"{u}hle, and K. K\"{o}hler.
\newblock Electron g factor in quantum wells determined by spin quantum beats.
\newblock {\em Solid State Communications}, 93(4):313 -- 317, 1995.

\bibitem{Oestreich1995}
M. Oestreich and W. W. R\"{u}hle.
\newblock Temperature Dependence of the Electron Land\'{e} $\mathit{g}$ Factor in
  GaAs.
\newblock {\em Phys. Rev. Lett.}, 74:2315--2318, Mar 1995.

\bibitem{Li2011}
Bin Li and F. M. Peeters.
\newblock Tunable optical Aharonov-Bohm effect in a semiconductor quantum ring.
\newblock {\em Phys. Rev. B}, 83:115448, Mar 2011.

\bibitem{Jeune1997}
P. Le Jeune, D. Robart, X. Marie, T. Amand, M. Brousseau, J. Barrau, V. Kalevich, and
  D. Rodichev.
\newblock Anisotropy of the electron Land\'{e} g factor in quantum wells.
\newblock {\em Semiconductor Science and Technology}, 12(4):380, 1997.
\end{references}


\end{document}